\documentclass{appolb}
\usepackage{graphicx}
\usepackage{amsmath,amssymb}
\usepackage[dvipsnames]{xcolor}
\usepackage{physics}

\usepackage{mciteplus}

\newcommand{\by}{\mathbf{y}}

\newcommand{\xpom}{x_\mathbb{P}}

\newcommand{\gev}{\mathrm{GeV}}

\newcommand{\rt}{{\mathbf{r}_\perp}}

\newcommand{\bt}{{\mathbf{b}_\perp}}
\newcommand{\bti}{{\mathbf{b}_{\perp,i}}}
\newcommand{\Deltat}{{\boldsymbol{\Delta}_\perp}}

\newcommand{\jpsi}{$\mathrm{J}/\psi$ }

\newcommand{\btheta}{{\boldsymbol{\theta}}}
\newcommand{\CalP}{{\mathcal{P}}}

\usepackage[breaklinks,colorlinks,citecolor=citcolor,urlcolor=blue,linkcolor=lcolor]{hyperref}
\definecolor{lcolor}{rgb}{0.5,0,0}
\definecolor{citcolor}{rgb}{0,0.3,0.0}

\begin{document}
\title{Bayesian inference of the fluctuating proton shape in DIS and hadronic collisions %
\thanks{Presented at XXIXth International Conference on Ultra-relativistic Nucleus-Nucleus Collisions (QM2022)}%
}
\author{Heikki Mäntysaari
\address{Department of Physics, University of Jyv\"askyl\"a, P.O. Box 35, 40014 University of Jyv\"askyl\"a, Finland, and \\
Helsinki Institute of Physics, P.O. Box 64, 00014 University of Helsinki, Finland}
\\[3mm]
Björn Schenke
\address{Physics Department, Brookhaven National Laboratory, Upton, NY 11973, USA}
\\[3mm]
Chun Shen
\address{Department of Physics and Astronomy, Wayne State University, Detroit, Michigan 48201, USA, and \\
RIKEN BNL Research Center, Brookhaven National Laboratory, Upton, NY 11973, USA}
\\[3mm]
Wenbin Zhao
\address{Department of Physics and Astronomy, Wayne State University, Detroit, Michigan 48201, USA}
}

\maketitle
\begin{abstract}
We determine the likelihood distribution for the model parameters describing the event-by-event fluctuating proton geometry at small $x$ by performing a Bayesian analysis within the Color Glass Condensate framework. The exclusive \jpsi production data from HERA is found to constrain the model parameters well, and we demonstrate that complementary constraints can be obtained from simulations of Pb+Pb collisions at the LHC.
\end{abstract}
  
\section{Introduction}

Determining the partonic structure of protons and nuclei is one of the main goals of the future nuclear DIS facilities such as the Electron-Ion Collider~\cite{AbdulKhalek:2021gbh}. Determining the spatial distribution of gluonic matter at small-$x$ is fundamentally interesting, and is also extremely important to provide input to simulations of heavy-ion collisions at RHIC and at the LHC where the final state hydrodynamical evolution transforms the initial coordinate space anisotropies into momentum space correlations.

In Deep Inelastic Scattering the simple structure of the photon probe allows for a precise determination of the gluonic structure of protons and nuclei. Exclusive processes such as the \jpsi production are especially important, as only in exclusive scattering it is possible to determine the total momentum transfer to the target hadron, which by definition is the Fourier conjugate to the impact parameter and as such provides access to the target geometry. In addition to the average geometry, it is important to understand the shape fluctuations that can be expected to play a major role when looking at fluctuation-dominated flow observables, e.g.~in proton-lead collisions at the LHC.

In our recent Letter~\cite{Mantysaari:2022ffw} we have performed a Bayesian analysis to extract the likelihood distribution for the model parameters describing the event-by-event fluctuating proton shape from HERA \jpsi production data~\cite{Alexa:2013xxa}, assuming that the nucleon substructure can be described in terms of gluonic hot spots as suggested in Refs.~\cite{Mantysaari:2016ykx}, see also Ref.~\cite{Mantysaari:2020axf} for a review. 


\section{Constraining model parameters using HERA data}

The scattering amplitude for exclusive vector meson production in the dipole picture can be written as~\cite{Kowalski:2006hc}
\begin{multline}
\label{eq:jpsi_amp}
    \mathcal{A}^{\gamma^*+p \to V + p} = 2i\int \dd[2]{\rt} \dd[2]{\bt}  \frac{\dd{z}}{4\pi} e^{-i \left[\bt - \left(\frac{1}{2}-z\right)\rt\right]\cdot \Deltat}  \\
    \times [\Psi_V^* \Psi_\gamma](Q^2,\rt,z) N_\Omega(\rt,\bt,\xpom).
\end{multline}
Here $\Psi_\gamma$ is the photon light front wave function describing the $\gamma^* \to q\bar q$ splitting, $\Psi_V$ is the vector meson wave function for which we use the Boosted Gaussian parametrization~\cite{Kowalski:2006hc}, and $N_\Omega$ is the dipole-proton scattering amplitude where $\Omega$ refers to a particular proton configuration. The transverse size of the dipole is $\rt$ and the proton-to-dipole distance is $\bt$. The fraction of the photon plus momentum carried by the quark is denoted by $z$, and $\xpom$ is the fraction of the target longitudinal momentum (in the infinite momentum frame) transferred in the process.

The coherent cross section corresponding to the events where the proton remains intact reads
\begin{equation}
     \frac{\dd \sigma^{\gamma^* + p \to V + p}}{\dd |t|}  = \frac{1}{16\pi} \left|\left\langle \mathcal{A}^{\gamma^*+p \to V + p} \right\rangle_\Omega\right|^2,
\end{equation}
and is sensitive to the average dipole-proton interaction, and as such to the average geometry. Here $\langle \rangle_\Omega$ correspond to an average overt the target configurations. On the other hand, calculating the total diffractive cross section and subtracting the coherent contribution one obtains the incoherent cross section corresponding to events in which the proton breaks up:
\begin{equation}
     \frac{\dd \sigma^{\gamma^* + p \to V + p^*}}{\dd |t|}  = \frac{1}{16\pi} \left[
     \left\langle \left|\mathcal{A}^{\gamma^* + p \to V + p}\right|^2\right \rangle_\Omega 
     - \left|\left\langle \mathcal{A}^{\gamma^*+p \to V + p} \right\rangle_\Omega\right|^2 \right]\,.
\end{equation}
As a variance, the incoherent cross section is sensitive to the amount of fluctuations at distance scale $\sim 1/\sqrt{|t|}$ in the scattering amplitude.

The dipole-proton scattering amplitude $N$ is obtained using the same approach as in the IP-Glasma calculation of the initial conditions for heavy ion collisions~\cite{Schenke:2012wb}, following Ref.~\cite{Mantysaari:2016ykx}. The local color charge density is assumed to be proportional to the local saturation scale $Q_s^2(\bt)$ extracted from the IPsat parametrization, and as such on the local density $T_p(\bt)$. 
We introduce an event-by-event fluctuating density by writing the density profile following Ref.~\cite{Mantysaari:2016ykx} as:
\begin{equation}
\label{eq:Tpfluct}
    T_p(\bt) = \frac{1}{N_q} \sum_{i=1 }^{N_q} p_i T_q(\bt-\bti), \quad  T_q(\bt) = \frac{1}{2\pi B_q} e^{-{\mathbf b}_\perp^2/(2B_q)}\,,
\end{equation}
and the coefficient $p_i$ allows for different normalizations for individual hot spots. This coefficient is sampled from a log-normal distribution whose width $\sigma$ is taken to be a model parameter, as well as the hot spot size $B_q$ and the proton size $B_p$, which is the width of a Gaussian probability distribution for the hot spot positions $\bti$. We additionally include repulsive short-range correlations between the hot spots by introducing a parameter $d_{q,\text{min}}$ which is the smallest allowed distance between the hot spots. The remaining model parameters are an infrared regulator $m$, the ratio between the local color charge density and the saturation scale  ($Q_s/(g^2\mu)$) that controls the overall normalization, and the number of hot spots $N_q$.

To determine the likelihood distribution for the model parameters we employ Bayesian Inference. It is a general and systematic method to constrain the probability distribution of model parameters $\btheta$ by comparing model calculations $\by(\btheta)$ with experimental measurements $\by_\mathrm{exp}$~\cite{sivia2006data} (\jpsi production data at $W=75\,\gev$ measured by H1~\cite{Alexa:2013xxa}).  According to Bayes' theorem the posterior distribution of model parameters satisfies
\begin{equation}
    \CalP(\btheta \vert \by_\mathrm{exp}) \propto \CalP(\by_\mathrm{exp} \vert \btheta) \CalP(\btheta).
\end{equation}
Here $\CalP(\by_\mathrm{exp} \vert \btheta)$ is the likelihood for model results with parameter $\btheta$ to agree with the experimental data that we calculate using Gaussian process emulators. The final posterior distribution is determined by using Markov Chain Monte Carlo sampling. For more details, see Ref.~\cite{Mantysaari:2022ffw} and references therein.

\section{Results}
\begin{figure}[tb]
\centerline{%
\includegraphics[width=10.8cm]{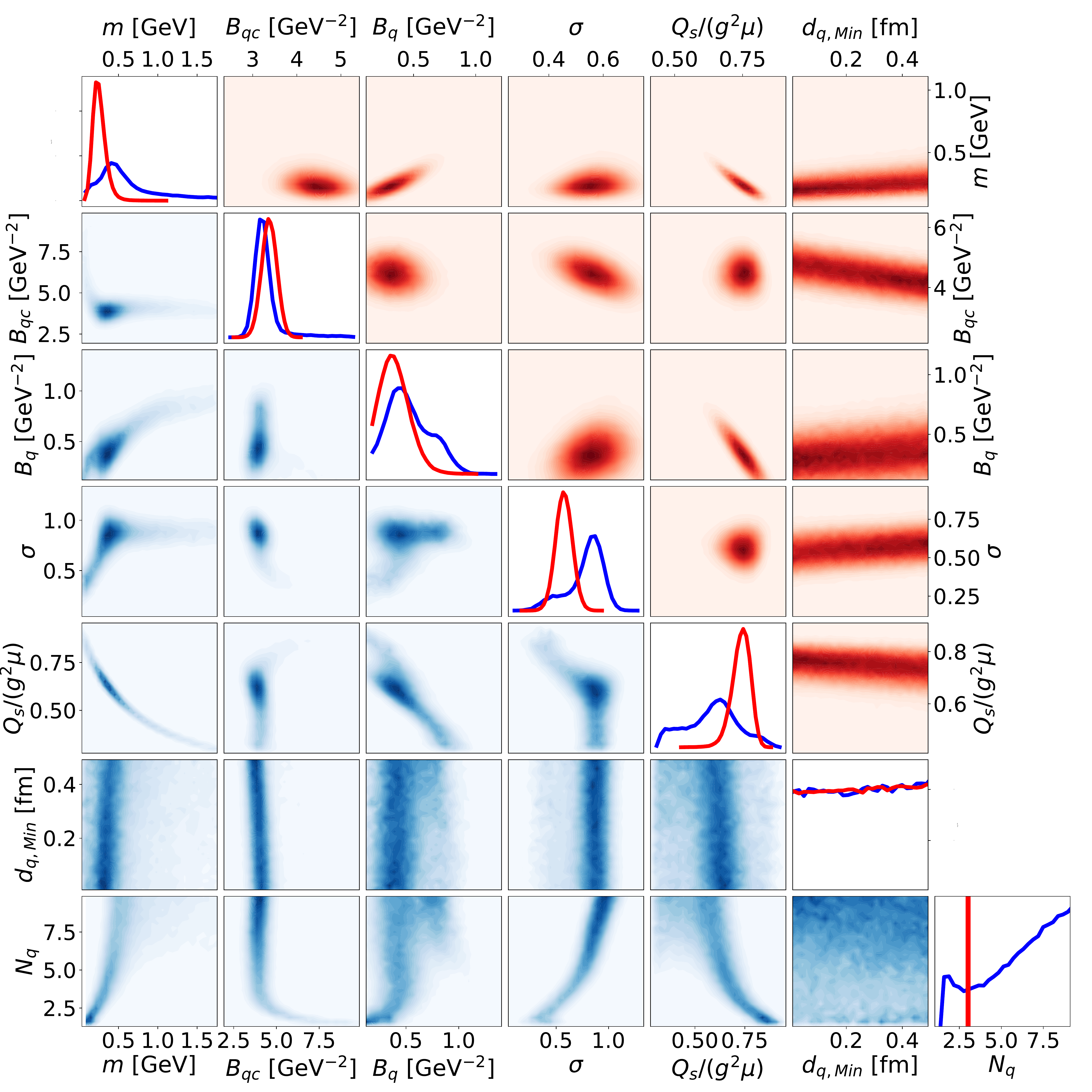}}
\caption{Posterior distribution of the model parameters. }
\label{Fig:posterior}
\end{figure}

The determined posterior distribution of model parameters is shown in Fig.~\ref{Fig:posterior}. The HERA data used to constrain the parameters corresponds to $\xpom \approx 10^{-3}$. We show separately results from two analyses, one with a fixed number of hot spots $N_q=3$ with results shown in red and in the upper right corner, the second with $N_q$ a free parameter, and results shown in blue and in the lower left corner. 

Most model parameters can be constrained well, except the parameter $d_{q,\text{min}}$ describing the repulsive short-range correlations between the hot spots that were found in Ref.~\cite{Albacete:2017ajt} to be necessary to describe high-multiplicity proton-proton collisions. This means that the \jpsi production data form HERA allows but does not require such repulsive correlations. Similarly the number of hot spots is not constrained by the data. This can be understood by noticing that there is a strong positive correlation between the number of hot spots $N_q$ and the hot spot density fluctuations $\sigma$. With large $N_q$ there are also very large density fluctuations which means that only a few hot spots actually dominate. Additionally with large $N_q$ the hot spots start to overlap which further reduces the ``effective number of hot spots''.

With both variable $N_q$ and $N_q=3$ we get an equally good description of the HERA data, which implies that the HERA data alone does not completely constrain the fluctuating geometry. Additional constraints can be obtained from heavy ion collisions. As a proof-of-concept, we take maximum likelihood parametrizations with $N_q=3$ and $N_q=9$, and use those to construct an initial condition for Pb+Pb collisions at $\sqrt{s}=5.02$ TeV. The initial condition and early evolution before the QGP phase is described using the IP-Glasma framework~\cite{Schenke:2012wb}. It is then coupled to MUSIC~\cite{Schenke:2010nt} hydrodynamical simulations of the plasma evolution, and to the UrQMD afterburner describing the more dilute hadronic phase~\cite{Bleicher:1999xi} (see \cite{Schenke:2020mbo} for a description of the entire framework).

\begin{figure}[h!]
\begin{minipage}{.49\textwidth}
\centering
		\includegraphics[width=0.9\textwidth]{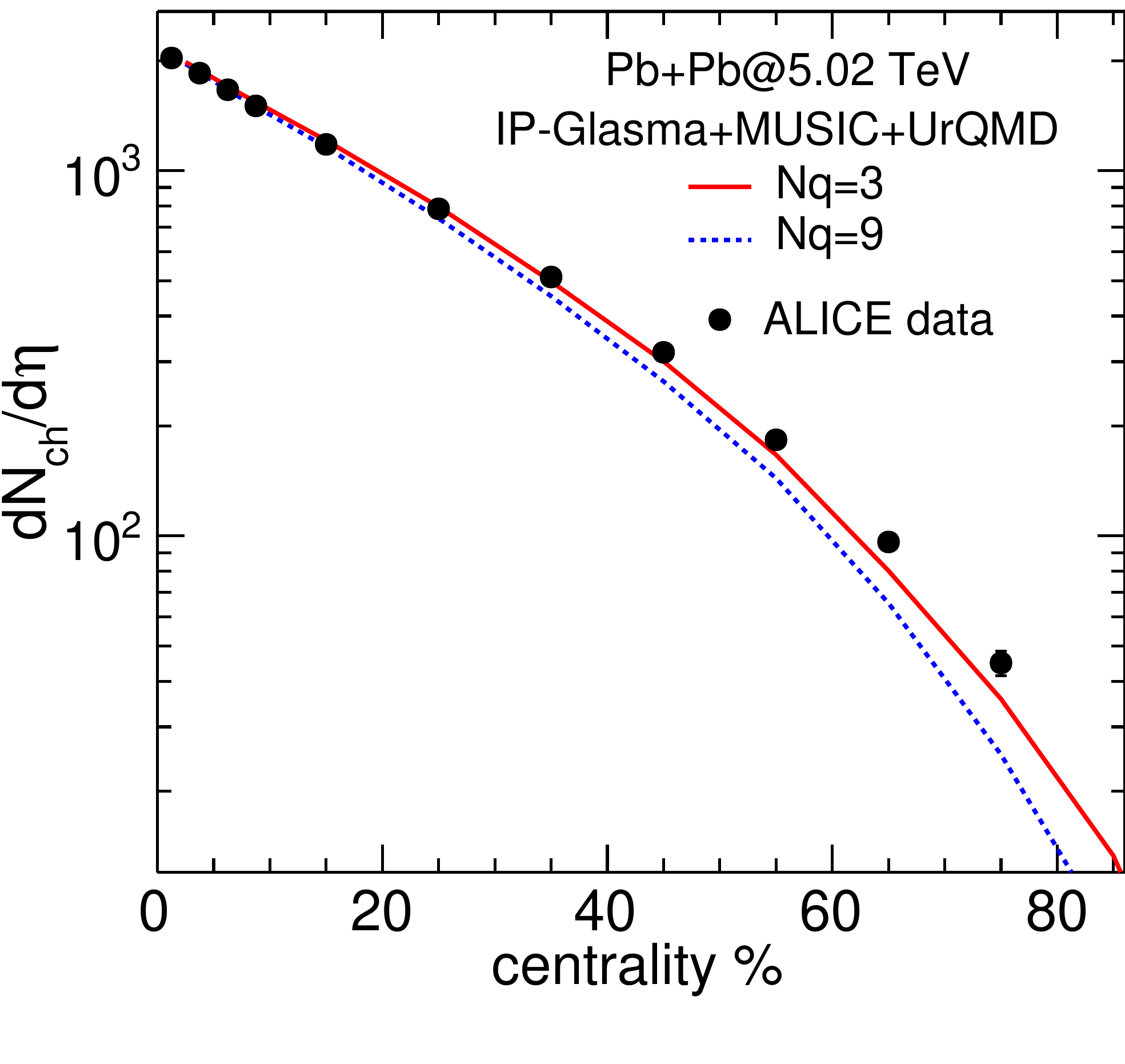}  
				\caption{Multiplicity distribution in Pb+Pb compared to ALICE data. }  
			\label{fig:pbpb_nch}

\end{minipage}
\quad
\begin{minipage}{.49\textwidth}
\centering
\includegraphics[width=0.9\textwidth]{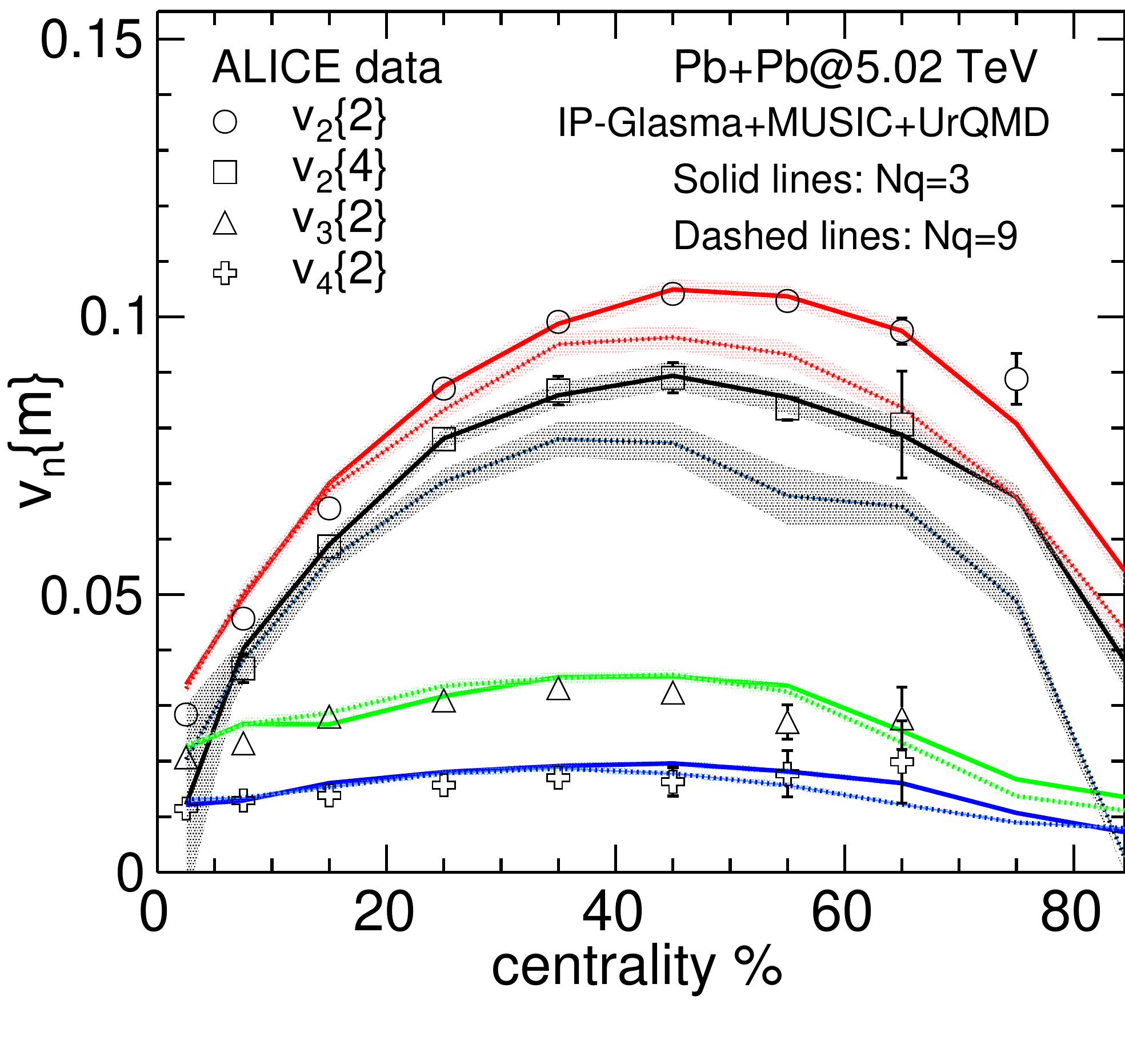} 
				\caption{Flow harmonics in Pb+Pb collisions compared to ALICE data. }  
			\label{fig:vnpbpb}
\end{minipage}
\end{figure}

The multiplicity distribution in Pb+Pb collisions is shown in Fig.~\ref{fig:pbpb_nch}. Note that the multiplicities in the most central bin match by construction. We find that the ALICE data {\bf cite} prefers the $N_q=3$ parametrization. Similarly the centrality dependence of the flow harmonics $v_2\{2\}$ and $v_2\{4\}$ prefer this parametrization. These results clearly indicate that the LHC data can provide further constraints on the fluctuating shape of the nucleons.

\section{Conclusions}

We have performed a  Bayesian analysis to extract the posterior likelihood distribution for the non-perturbative parameters describing the event-by-event fluctuating proton geometry using the HERA \jpsi production data. Most of the model parameters are well constrained by the data, except that the potential repulsive short-range correlations can not be determined from this data. The obtained likelihood distribution can be used to systematically take into account uncertainties in the proton geometry when calculating any other observable that depends on the event-by-event fluctuating geometry.
We have further demonstrated that complementary constraints can be obtained from simulations of heavy ion collisions where the initial nucleon geometry affects the space-time evolution of the produced QGP. 

\subsection*{Acknowledgments}
B.P.S. and C.S. are supported under DOE Contract No.~DE-SC0012704 and Award No.~DE-SC0021969, respectively.  C.S. acknowledges a DOE Office of Science Early Career Award.
H.M. is supported by the Academy of Finland, the Centre of Excellence in Quark Matter, and projects 338263 and 346567. W.B.Z. is supported by the National Science Foundation (NSF) under grant numbers ACI-2004571.

\bibliographystyle{JHEP-2modM}
\bibliography{refs}

\end{document}